\documentclass[conference]{IEEEtran}
\IEEEoverridecommandlockouts

\usepackage{graphicx}
\usepackage{url}
\usepackage{booktabs}
\usepackage{tabularx}

\usepackage[colorinlistoftodos]{todonotes}
\usepackage{paralist}
\usepackage{comment}
\usepackage{amsmath}
\usepackage{amssymb}
\usepackage{amssymb}
\usepackage{tikz}
\usepackage{pgfplots}
\usepgfplotslibrary{groupplots}
\usepackage{cite}

\begin{document}
\title{MEC-enhanced Information Freshness for Safety-critical C-V2X Communications}
\author{\IEEEauthorblockN{Mustafa Emara, Miltiades C. Filippou, Dario Sabella}
\IEEEauthorblockA{Next Generation and Standards, Intel Deutschland GmbH, Neubiberg, Germany \\
Email:$\{$first name.last name$\}$@intel.com}
}

\maketitle
\IEEEpeerreviewmaketitle

\begin{abstract}
	Information freshness is a status update timeliness indicator of utmost importance to several real-time applications, such as connected and autonomous driving. The Age-of-Information (AoI) metric is widely considered as useful to quantify the information freshness of delivered messages to the involved entities. Recently, the advent of Multi-access Edge Computing (MEC) promises several performance benefits for Vehicular-to-Everything (V2X) communications, emphasizing on the experienced End-to-End (E2E) message delay. In this paper, we argue that, when it comes to safety-critical use cases, such as the one of Vulnerable Road User (VRU), additional metrics can be more insightful to evaluate and address scalability issues in dense urban environments. In particular, the impact of the packet inter-arrival time on the timeliness of VRU messages arriving at nearby vehicles can be directly assessed by exploiting the AoI metric. For that purpose, assuming a MEC-enabled multi-VRU system setting, we model the AoI and, by means of a performance comparison to the state-of-the-art network architecture based on numerical evaluations, we provide evidence of the information freshness and system scalability enhancements offered by MEC infrastructure deployment for different system parameter settings involving a large number of connected entities.
\end{abstract}
\section{Introduction}
\label{sec:introduction}
\subsection{MEC-enabled C-V2X communications}
Vehicle-to-Everything (V2X) communication technology aims to provide radically improved road safety and driving experience via reliable and low-latency wireless services \cite{5GAA}. Efficient V2X system operation is based on a large set of sensors such as cameras, Light Detection and Ranging (LiDAR) sensors and radars providing an enhanced environmental awareness to vehicles, pedestrians and road infrastructure through the exchange of critical messages among connected entities \cite{Schiegg2019}. Information links may be established either via short range connections, or assisted by the cellular network (i.e., cellular-V2X (C-V2X) communication), or through a combination of both technologies \cite{Shen2018}. 

With regards to the C-V2X technology, traditional approaches involving communication through remote cloud servers, are expected to significantly limit the support of delay-critical V2X services, as the End-to-End (E2E) delay between message transmission and reception among roadside connected entities is affected by the introduced backhaul delays, together with the ones introduced by both the Core Network (CN), as well as the Transport Network (TN). Such delay bottlenecks will be even more notable when it comes to dense deployments of connected entities (e.g., vehicles, pedestrians). To alleviate these performance limitations, operators are currently expressing growing interest in the use of Multi-access Edge Computing (MEC), which allows applications to be instantiated at the edge of the network, and, hence, provides a low-latency environment, due to close proximity to user terminals. When it comes to the automotive domain, MEC technology has been shown to provide performance gains for various V2X system setups \cite{Zhou2018,Campolo2019,Emara2018}. Hence, the automotive industry is expected to significantly benefit from the deployment of MEC infrastructure in C-V2X systems.

\subsection{The VRU use case and its evaluation metrics}
The 5G Automotive Association (5GAA) has taken into account the emergence of a plurality of new, innovative use cases and, therefore, has identified the following seven C-V2X use case groups: \begin{inparaenum}[(a)] \item safety, \item vehicle operations management, \item convenience, \item autonomous driving, \item platooning, \item traffic efficiency and environmental friendliness, as well as \item society and community \end{inparaenum} \cite{5GAA2}. Focusing on the safety use case group, \emph{Vulnerable Road User} (VRU) is about the safe interaction between vehicles and non-vehicle road users (pedestrians, motorbikes, etc.) via the exchange of periodic Cooperative Awareness Messages (CAM) \cite{Sabella2017}.

The VRU use case introduced by 5GAA incorporate time-critical scenarios such as: \begin{inparaenum}[(i)] \item the awareness of the presence of VRUs near potentially dangerous situations and \item collision risk warning \end{inparaenum} \cite{5GAA2}. For the both VRU cases, a straightforward performance metric to evaluate technology solutions targeting the efficient operation of such scenarios is the experienced E2E signaling latency between connected entities such as a VRU and approaching vehicles. Towards this end, the objective of \cite{Emara2018} was, focusing on a freeway VRU scenario, to evaluate the E2E latency performance achieved through the collocated deployment of MEC hosts and cellular network Evolved Node Bs (eNBs) and compare it to the one of the state-of-the-art cellular network, where packet processing takes place in the remote cloud. According to the presented numerical evaluation results, it was evident that the MEC-based system architecture outperformed its cloud-based counterpart for a number of system setups. 

Nevertheless, the CAM message E2E delay metric, although useful, it is insufficient to fully characterize system performance, as it overlooks the impact of the CAM sampling period (equivalently, the packet inter-arrival time). According to ETSI TR 103 300-1 \cite{etsi.its.103300-1}, it is exactly the periodicity of broadcast messages \emph{together} with the communication latency that contributes to the age of data elements, as, the latter may affect e.g., the consistency between the VRU positioning accuracy and the received positioning data elements evolution. 

\subsection{Our contributions}
Inspired by the gaps identified above, the goal of this paper is to evaluate the AoI performance for the VRU case with respect to a system setup consisting of multiple VRUs, vehicles, radio nodes and MEC infrastructure collocated to the Radio Access Network (RAN) nodes. To the best of our knowledge, such a performance study has not been undertaken so far, as in technical works, such as \cite{Ni2018}, although broadcast messages are scheduled per a sum AoI minimization criterion, the information freshness performance of a MEC-enabled system for a VRU scenario is not calculated at all.

The remainder of this paper is organized as follows: in Section \ref{sec:sys_model}, we present an overview of the studied system setup, along with details of the VRU scenario. Section \ref{sec:aoi_model} provides a description of the AoI metric for the focused scenario and clarifies upon how AoI depends on the E2E signaling latency. Section \ref{sec:num_evaluation} presents and discusses numerical evaluation results, while Section \ref{sec:conclusion} draws some conclusions of the paper.
\section{System Model}
\label{sec:sys_model}
In what follows in this section we provide a detailed description of the investigated VRU system setup, we also explain VRU message modeling and transmission per the use case description in \cite[Section 4.12]{5GAA2}, concentrating on the scenario of the awareness of the presence of VRUs near potentially dangerous situations and we also clarify upon the assumed physical layer parameter values. 

\subsection{VRU system scenario}
\begin{figure}[t]
	\includegraphics[width=0.48\textwidth]{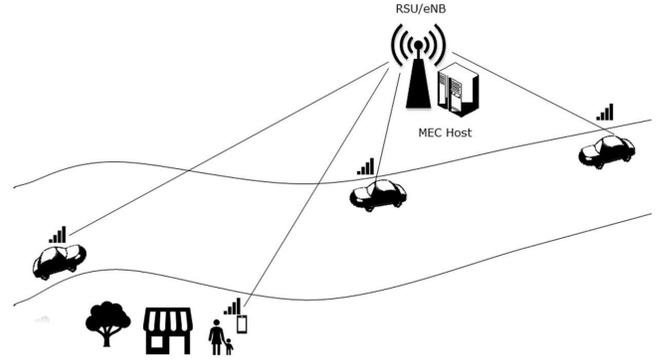}
	\caption{The investigated VRU system setup.}
	\label{fig:sys_model}
\end{figure}

The system setup is depicted in Fig.~\ref{fig:sys_model}. We  assume a freeway road environment consisting of one lane per direction under the coverage of an eNB or a Roadside Unit (RSU) collocated with a MEC host of given processing capabilities. We choose to focus on a specific road segment under the coverage of a single radio access point, as the investigation of the impact of radio handovers on VRU message freshness is beyond the scope of the evaluation conducted in this paper and is left for future work.
With regards to the assumed road environment, the speed of each vehicle entering the road area under cellular coverage is drawn by a uniformly distributed random variable (i.e., $ \text{v} \in \mathcal{U}(\text{v}_{min}, \text{v}_{max})$). 

At the same time, a number of VRUs, such as pedestrians, cyclists and other connected entities of lower mobility are located on a pedestrian area; such a populated area can be mapped to real-world scenarios like gas stations or other service points across a freeway. According to the focused scenario, each of the existent VRUs, which operates its own User Equipment (UE), periodically informs a specific cluster of approaching vehicles of its presence by means of notifications providing location information, among possible other updates. The messages originating from the VRUs are received by the radio infrastructure node in the uplink using the Uu radio interface and are subsequently processed at the MEC host. The processed messages are broadcasted to the intended cluster of approaching vehicles under cellular coverage using the downlink channel, in order to inform vehicle drivers, who can take appropriate actions to avoid dangerous situations.

\subsection{VRU message modeling}
\label{sec:message_model}
The VRU awareness notification information aiming at drivers of approaching vehicles is packed up in periodically transmitted VRU messages \cite[Section 4]{etsi.its.103300-1}. VRU messages are useful together with CAM messages which are exchanged within the Intelligent Transportation System (ITS) to create and maintain awareness of the network and to support cooperative performance of vehicles using the road network. VRU messaging is especially useful for safety-related applications, where it can be exploited, e.g., for crash prevention purposes \cite{Lyamin2018}. Accordingly, a proper VRU periodic messaging model needs to be adopted to provide sensible insights on the VRU use case. Throughout this work, assuming the existence of $K$ VRUs over the focused area under cellular coverage, we consider a network-wide homogeneous asynchronous VRU signal traffic model. In particular, let the $k$-th VRU generate data packets of size of $l_k \in \mathcal{U}(l_\text{min},l_\text{max})$ bits at random starting time offsets, denoted as $\beta_k, \; \forall k = 1,2,\cdots,K$. A new VRU message is generated periodically at the time slot corresponding to $\beta_k + nT; \; \forall n = 1,2,\cdots$, where, $T$ stands for the VRU messaging time period. Due to the shared nature of the wireless channel, the assigned time offsets for the VRUs dictate the number of VRUs simultaneously requesting access to the channel which, in its turn, affects the number of available uplink radio resources per VRU. 

\subsection{Physical layer parameters}
All considered vehicles and VRUs are assumed to be served via the Uu interface by their serving eNB, based on the pathloss model adopted from the \textit{WINNER+} project \cite{WINNER2007}, as follows
\begin{align}
\text{PL (dB)} &= 22.7\text{log}_{10}(d) - 17.3\text{log}_{10}(\tilde{h}_{\text{eNB}}) -17.3\text{log}_{10}(\tilde{h}_{\text{UE}}) \nonumber\\
&+ 2.7\text{log}_{10}(f_\text{c}) - 7.56,
\end{align}
where $d$ is the transmitter-receiver distance, $f_\text{c}$ is the center carrier frequency and $\tilde{h}_{\text{eNB}}$ and $\tilde{h}_{\text{UE}}$ represent the effective antenna heights, respectively at the eNB and at the UE (operated by the VRU). The latter quantities are computed as follows: $\tilde{h}_{\text{eNB}} = h_{\text{eNB}} - 1.0$ and  $\tilde{h}_{\text{UE}} = h_{\text{UE}}- 1.0$, with $h_{\text{eNB}}$ and $h_{\text{UE}}$ being the actual antenna heights (i.e., in meters). Also, independent and identically distributed (i.i.d.) random variables are used to model the fast fading and shadowing attenuations. Finally, it should be noted that the packet scheduler employed in our work equally distributes the available radio resources over all scheduled VRUs and vehicles, as well as that no VRU message transmission failures occur.

After laying out the main aspects of the proposed framework, an AoI-based analysis for the examined VRU use case is presented, taking into account the proposed, MEC-assisted access network architecture.
\section{Age-of-Information and its dependency on E2E message latency}
\label{sec:aoi_model}

\subsection{Quantifying the freshness of VRU messages}
Concentrating on the VRU use case, \cite{Emara2018} showcased the E2E latency-related benefits of introducing MEC system deployment over a state-of-the-art cellular network. As it will be explained in further detail in this section, we argue that, apart from the E2E latency, the freshness of continuous updates of nodes within a V2X system is another fundamental performance indicator to ensure efficient service functionality, especially for safety-critical situations. This implies continuous information update about the real-time state between a given source and its targeted destination \cite{NGMNA2016}. The AoI metric proposed in \cite{Kaul2012} characterizes the freshness of information at the receiver and has recently received increased attention as it is a useful metric to evaluate the efficiency of technology solutions for various vertical industries, such as the automotive one. The AoI at a given time stamp (i.e., observation point) is defined as the current time stamp minus the time at which the observed state (or packet) was generated \cite{Yates2019}. 
\begin{figure}[t]
	\includegraphics[width=0.48\textwidth]{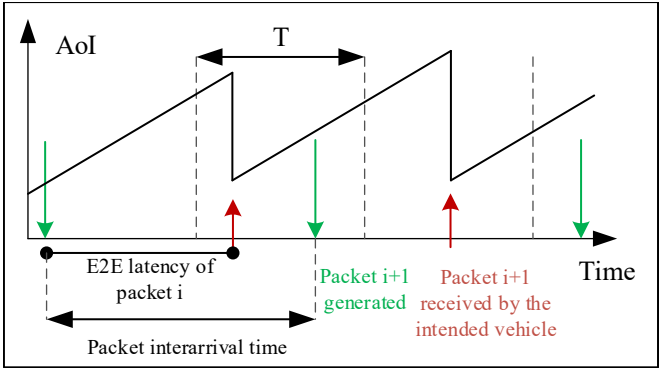}
	\caption{AoI evolution over time for a given VRU signaling source and a specific vehicle (cluster member).}
	\label{fig:AoI_model}
\end{figure}

In contrast to solutions involving traditional time-centric requirements, such as delay and jitter, the design of an AoI-minimizing status update signaling policy can enhance the timeliness of such updates in a way the traditional metrics cannot. The reason is that, per the definition of the AoI metric, the inter-arrival time of generated VRU packets may significantly impact the AoI and, hence, the overall system timeliness performance. Consequently, for the examined use case, to ensure an -almost- real-time VRU awareness across the vehicles, it is the timeliness of VRU messages received by nearby vehicles that would rather need to be monitored and improved, e.g., by properly varying the VRU packet generation traffic. In relation to that, a critical challenge is how to maintain timely VRU status updates across all approaching connected vehicles \cite{Jiang2019}. 

For the considered vehicular time-slotted system, the AoI function, $\Delta_{k}(t)$, tracks the AoI evolution over time, $t$, at each of the cluster member vehicles aimed to be reached by the $k$-th VRU. Let $G_{k}(t)$ denote the packet generation time stamp for the $k$-th VRU; then, focusing on a specific vehicle/ cluster member, the AoI at the $(t+1)$-st time slot, denoted by $\Delta_{k}(t+1)$, is computed recursively as follows
\begin{equation}
\Delta_{k}(t+1)=\left\{\begin{array}{ll}{\Delta_{k}(t)+1,} & {\text {if no update was received, }} \\ {t-G_{k}(t)+1,} & {\text { otherwise. }}\end{array}\right.
\end{equation}
A visualization of the temporal evolution of the AoI is depicted in Fig. \ref{fig:AoI_model}, where, one can observe how the AoI evolves linearly with time till a new VRU message is successfully received by an intended vehicle. In this work, focusing on a given VRU, we consider the cluster-wide \emph{peak AoI} (PAoI), which is defined as the AoI observed at the farthest member of the vehicle cluster targeted by the VRU, when achieved immediately before this vehicle receives a new VRU message \cite{Huang2015}. The PAoI represents the temporally averaged peaks attained by the AoI function shown in Fig. \ref{fig:AoI_model}.  As the PAoI provides insights on \emph{guaranteed} system performance, we deem it as an important metric for the investigated VRU scenario. Mathematically, the PAoI of the $k$-th VRU, when averaged over time, is 
\begin{equation}\label{eq:peakAoI}
\Delta_k^\text{p} = \mathbb{E}_{t}\Big\{\mathcal{I} + \mathcal{T} \Big\},
\end{equation}
where $\mathbb{E}_{t}\{.\}$ is the temporal expectation operator, while, $\mathcal{I}$ and $\mathcal{T}$ denote the inter-arrival time between consecutive VRU messages and the E2E latency of a given VRU message, respectively. Based on the periodic nature of the VRU messages described in Section \ref{sec:message_model}, eq.~(\ref{eq:peakAoI}) can be rewritten as follows 
\begin{equation}
\Delta_k^\text{p}\ = T + \mathbb{E}_{t}\Big\{ \mathcal{T} \Big\}.
\end{equation}

\subsection{AoI modeling for different network architectures}
As highlighted earlier, the objective of this work is to investigate the VRU awareness timeliness performance achieved through collocated deployment of a MEC and cellular network infrastructure and compare it to the one of conventional cellular system architecture incorporating (distant) cloud infrastructure. To accomplish this aim, in this section we model the various latency components corresponding to VRU packet transmission, routing and processing for both the proposed and conventional system approaches.

Regarding the conventional cellular network architecture approach, the one-way VRU messaging latency is modeled as $\mathcal{T}_\text{one-way} = T_{\text{UL}} + T_{\text{BH}} + T_{\text{TN}} + T_{\text{CN}} + T_{\text{Exc}}$, where $T_{\text{UL}}$ is the radio UL transmission latency, $T_{\text{BH}}$ is the backhaul network latency, $T_{\text{TN}}$ is the TN latency, $T_{\text{CN}}$ is the CN latency and $T_{\text{Exc}}$ is the VRU message processing latency. Consequently, the E2E latency for the conventional cellular architecture, is expressed as
\begin{equation}\label{eq:total_latency_conv}
\mathcal{T}_\text{E2E, C} = T_{\text{UL}} + \underbrace{2(T_{\text{BH}} + T_{\text{TN}} + T_{\text{CN}})}_\textrm{Network latency} + T_{\text{Exc}} +  T_{\text{DL}},
\end{equation}
where, $T_{\text{DL}}$ represents the downlink transmission latency.\footnote{Latency from the eNB to the MEC host and vice versa is not considered and left for future work.} For the proposed, MEC-enabled network approach, the network latency marked in eq.~(\ref{eq:total_latency_conv}) can be avoided via processing the VRU packets at the MEC host, collocated with the connected eNB, therefore, in this case, the E2E latency is given by
\begin{equation}\label{eq:total_latency}
\mathcal{T}_\text{E2E, MEC} = T_{\text{UL}} + T_{\text{Exc}} +  T_{\text{DL}}.
\end{equation}
For detailed information regarding the models adopted to evaluate the E2E latency components, the reader is kindly referred to \cite[Section III]{Emara2018}. Furthermore, considering the assumed model for VRU packet generation as well as the experienced E2E latency, the network-wide PAoI, averaged over all $K$ VRUs in the network is evaluated as 
\begin{equation}\label{eq:network_PAoI}
\tilde{\Delta}_j^\text{p} = \mathbb{E}_k \{ \Delta_k^\text{p} \} = \frac{1}{K}\sum_{i=1}^{K} (T+\mathcal{T}_\text{E2E, j}),
\end{equation}
where $j\in\{\text{C, MEC}\}$. 

\section{Numerical evaluation}
\label{sec:num_evaluation}

To evaluate the effect of MEC infrastructure deployment on the information freshness performance for the VRU use case of C-V2X communications, we consider different simulation scenarios by varying the values of two main system parameters: i) the VRU spatial density, which sheds light on the achieved system scalability, and ii) the VRU message frequency, which defines how often a VRU generates a packet. For both the proposed and conventional cellular network architectures, the metric of interest is the network-wide PAoI, mathematically defined in eq.~\eqref{eq:network_PAoI}. The values of all parameters involved in this simulations campaign are based on \cite[Table I]{Emara2018}, unless otherwise stated. 

\subsection{Impact of VRU density}

\begin{figure}
	\centering
	\vspace{0.3cm}
	\includegraphics[width=0.48\textwidth]{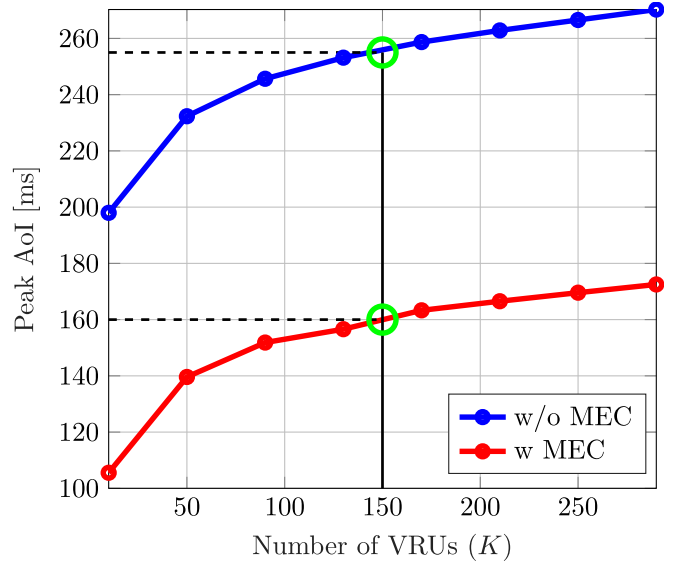}
	\caption{Network-wide PAoI with increasing VRU density for  $T=100$ ms.} 
	\label{fig:PAoI_incVRU}
\end{figure}

We first investigate the network-wide PAoI behavior of the system for increasing VRU density, assuming a given geographical area (i.e., a roadside service point). Due to the periodic nature of VRU message generation, each VRU UE is set to transmit its packet every $T$ milliseconds, on average. As a result, for an increased number of VRUs, the generated VRU message traffic per unit time within the network will increase as well, hence, resulting to less radio and processing resources allocated per VRU to transmit and process each VRU message, respectively. In Fig.~\ref{fig:PAoI_incVRU}, assuming that $T$=100 milliseconds, the network-wide  PAoI performance is illustrated, for both the MEC-enabled and conventional network architecture variants. Clearly, for all considered values of $K$, MEC infrastructure utilization provides a lower PAoI, thus, higher information timeliness, which, in its turn, is translated into better VRU awareness, compared to the conventional cellular architecture. As an example, for $K=150$ VRUs, the achieved PAoI is equal to $\tilde{\Delta}_{\text{MEC}}^\text{p}=160$ milliseconds, which is only a fraction of $\tilde{\Delta}_{\text{C}}^\text{p}=258$ milliseconds achieved by the conventional network architecture. Such a, nearly $61\%$, reduction in PAoI, is due to the exploitation of processing resource proximity offered by the deployed MEC host. Also, as expected, for both system architecture variants, we observe a monotonically increasing behavior of the PAoI as a function of the VRU load, owing to the increasing demand for radio and processing resources.

\subsection{Impact of VRU packet inter-arrival time}

\begin{figure}
	\centering
	\includegraphics[width=0.48\textwidth]{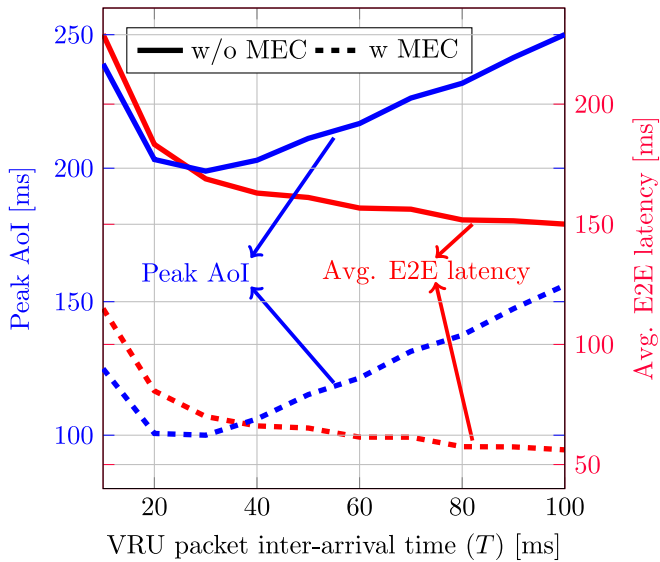}
	\caption{Peak AoI and average E2E latency for increasing VRU packet inter-arrical time with $K = 100$ VRUs.} 
	\label{fig:PAoI_incMsgFreq}
\end{figure}

To jointly evaluate the effect of VRU packet generation periodicity on system-wide timeliness and E2E delay performance, along with the performance gains provided by the existence of MEC infrastructure, assuming the existence of $K$=100 VRUs in the system, we measure the network-wide PAoI together with the average E2E VRU message latency for various VRU packet inter-arrival times, $T \in [10 \textrm{ms}, 100 \textrm{ms}]$. Fig.~\ref{fig:PAoI_incMsgFreq} depicts the numerical evaluation results, where, PAoI and average E2E delay values appear in the left and right hand side vertical axes of the figure, respectively. Apart from the clear performance gains when introducing a MEC host collocated with the cellular radio access node, one can identify two different performance behaviors with respect to the VRU packet inter-arrival time for both network architecture options. When $T \in [10 \textrm{ms}, 30 \textrm{ms}]$, both the achieved PAoI and the average E2E latency performance curves are monotonically decreasing, as a function of $T$. Such a behavior is justified as, in this regime, in contrast to $T$, the average E2E delay, which dominantly contributes to the PAoI, progressively reduces due to the reducing congestion on the available resources; this PAoI regime can be labeled as a \emph{resource stagnation-driven} one. On the contrary, when $T \in [30 \textrm{ms}, 100 \textrm{ms}]$, it is observed that, although the average E2E latency continues to decrease, as a function of $T$, the achieved PAoI starts to increase. This behavior differentiation occurs, because, focusing on the E2E latency, the resource contention among the VRUs radically decreases, as the set of possible VRU transmission offsets becomes fairly larger, hence, leading to lower overall delay per VRU message. Nevertheless, larger values of $T$ imply less frequent VRU status updates, resulting to higher values of the PAoI, as $T$ now decisively contributes to it; this PAoI regime can be labeled as an \emph{update scarcity-driven} one. In summary, we observe the limitations of considering the E2E latency as the sole objective of system design, with regards to time-critical applications for C-V2X communications. To alleviate these limitations, AoI minimization shall be the overall design objective when it comes to such applications and use cases.

\section{Conclusion}
\label{sec:conclusion}

In this work, focusing on the AoI as a means to quantify the information freshness of VRU messages, we have proposed a cellular network architecture encompassing MEC infrastructure. By numerically evaluating the achieved PAoI for both the proposed, MEC-assisted and the state-of-the-art network architectures, we have provided evidence of the VRU scalability enhancements provided by the deployment of roadside MEC infrastructure. In particular, we have shown that, for a given VRU load, the network-wide PAoI of the conventional system architecture can be reduced by nearly $61\%$ when a MEC-enabled network architecture is taken into account, instead. Also importantly, assuming a dense VRU setting, we have identified VRU packet inter-arrival time regimes, where, the PAoI is dominantly affected by either the encountered shortage of radio and processing resources, or, by the VRU message scarcity. Future works may include the possibility to assess instantaneous AoI behavior, by extending the models adopted for the involved network components, as well as the investigation of the feasibility of roadside safety constraints assuming a MEC-enabled architecture.

\bibliographystyle{IEEEtran}
\bibliography{bibliography}

\begin{thebibliography}{10}
\providecommand{\url}[1]{#1}
\csname url@samestyle\endcsname
\providecommand{\newblock}{\relax}
\providecommand{\bibinfo}[2]{#2}
\providecommand{\BIBentrySTDinterwordspacing}{\spaceskip=0pt\relax}
\providecommand{\BIBentryALTinterwordstretchfactor}{4}
\providecommand{\BIBentryALTinterwordspacing}{\spaceskip=\fontdimen2\font plus
\BIBentryALTinterwordstretchfactor\fontdimen3\font minus
  \fontdimen4\font\relax}
\providecommand{\BIBforeignlanguage}[2]{{%
\expandafter\ifx\csname l@#1\endcsname\relax
\typeout{** WARNING: IEEEtran.bst: No hyphenation pattern has been}%
\typeout{** loaded for the language `#1'. Using the pattern for}%
\typeout{** the default language instead.}%
\else
\language=\csname l@#1\endcsname
\fi
#2}}
\providecommand{\BIBdecl}{\relax}
\BIBdecl

\bibitem{5GAA}
5GAA, ``White paper on {C}-{V}2{X} - conclusions based on evaluation of
  available architectural options,'' 2019.

\bibitem{Schiegg2019}
F.~A. {Schiegg}, N.~{Brahmi}, and I.~{Llatser}, ``Analytical performance
  evaluation of the collective perception service in {C}-{V}2{X} mode 4
  networks,'' in \emph{2019 IEEE Intelligent Transportation Systems Conference
  (ITSC)}, Oct. 2019, pp. 181--188.

\bibitem{Shen2018}
X.~{Shen}, J.~{Li}, L.~{Chen}, J.~{Chen}, and S.~{He}, ``Heterogeneous
  {L}{T}{E}/{D}{S}{R}{C} approach to support real-time vehicular
  communications,'' in \emph{2018 10th International Conference on Advanced
  Infocomm Technology (ICAIT)}, Aug. 2018, pp. 122--127.

\bibitem{Zhou2018}
S.~{Zhou}, P.~P. {Netalkar}, Y.~{Chang}, Y.~{Xu}, and J.~{Chao}, ``The
  {M}{E}{C}-based architecture design for low-latency and fast hand-off
  vehicular networking,'' in \emph{2018 IEEE 88th Vehicular Technology
  Conference (VTC-Fall)}, Aug. 2018, pp. 1--7.

\bibitem{Campolo2019}
C.~{Campolo}, A.~{Iera}, A.~{Molinaro}, and G.~{Ruggeri}, ``{M}{E}{C} support
  for 5{G}-{V}2{X} use cases through {D}ocker containers,'' in \emph{2019 IEEE
  Wireless Communications and Networking Conference (WCNC)}, Apr. 2019, pp.
  1--6.

\bibitem{Emara2018}
M.~{Emara}, M.~C. {Filippou}, and D.~{Sabella}, ``{M}{E}{C}-assisted end-to-end
  latency evaluations for {C}-{V}2{X} communications,'' in \emph{2018 European
  Conference on Networks and Communications (EuCNC)}, Jun. 2018, pp. 1--9.

\bibitem{5GAA2}
5GAA, ``White paper on {C}-{V}2{X} use cases; methodology, examples and service
  level requirements,'' 2019.

\bibitem{Sabella2017}
{D. Sabella et al.}, ``Toward fully connected vehicles: Edge computing for
  advanced automotive communications,'' [Online]. Available:
  http://5gaa.org/news/toward-fully-connected-vehicles-edge-computing-for-advanced-automotive-communications/,
  Tech. Rep., 2017.

\bibitem{etsi.its.103300-1}
ETSI, ``{Intelligent Transport System (ITS); Vulnerable Road Users (VRU)
  awareness; Part 1: Use Cases definition; Release 2},'' European
  Telecommunications Standards Institute (ETSI), Technical Report (TR) ETSI TR
  103 300-1 v2.1.1, Sept. 2019.

\bibitem{Ni2018}
Y.~{Ni}, L.~{Cai}, and Y.~{Bo}, ``Vehicular beacon broadcast scheduling based
  on age of information ({A}o{I}),'' \emph{China Communications}, vol.~15,
  no.~7, pp. 67--76, Jul. 2018.

\bibitem{Lyamin2018}
N.~{Lyamin}, A.~{Vinel}, M.~{Jonsson}, and B.~{Bellalta}, ``Cooperative
  awareness in {V}{A}{N}{E}{T}s: On {E}{T}{S}{I} {E}{N} 302 637-2
  performance,'' \emph{IEEE Transactions on Vehicular Technology}, vol.~67,
  no.~1, pp. 17--28, Jan. 2018.

\bibitem{WINNER2007}
``{WINNER II} channel models, {D}1.1.2 v1.2,'' [Online]. Available:
  https://cept.org/files/8339/winner2\%20-\%20final\%20report.pdf, Tech. Rep.

\bibitem{NGMNA2016}
NGMNA, ``Recommendations for {NGMN KPIs} and requirements for 5{G},''
  \emph{Next Generation Mobile Networks Alliance}, 2016.

\bibitem{Kaul2012}
S.~{Kaul}, R.~{Yates}, and M.~{Gruteser}, ``Real-time status: How often should
  one update?'' in \emph{2012 Proceedings IEEE INFOCOM}, Mar. 2012, pp.
  2731--2735.

\bibitem{Yates2019}
R.~D. {Yates} and S.~K. {Kaul}, ``The age of information: Real-time status
  updating by multiple sources,'' \emph{IEEE Transactions on Information
  Theory}, vol.~65, no.~3, pp. 1807--1827, Mar. 2019.

\bibitem{Jiang2019}
Z.~{Jiang}, B.~{Krishnamachari}, X.~{Zheng}, S.~{Zhou}, and Z.~{Niu}, ``Timely
  status update in wireless uplinks: Analytical solutions with asymptotic
  optimality,'' \emph{IEEE Internet of Things Journal}, vol.~6, no.~2, pp.
  3885--3898, Apr. 2019.

\bibitem{Huang2015}
L.~{Huang} and E.~{Modiano}, ``Optimizing age-of-information in a multi-class
  queueing system,'' in \emph{2015 IEEE International Symposium on Information
  Theory (ISIT)}, Jun. 2015, pp. 1681--1685.

\end{thebibliography}

\end{document}